\documentclass[aps,prl,twocolumn,superscriptaddress]{revtex4}
\usepackage{graphicx}
\usepackage[latin1]{inputenc}
\usepackage{textcomp}
\usepackage{mathptmx}
\usepackage[usenames]{color}

\begin{document}

\title{Ultra-cold Single-Atom Quantum Heat Engines}

\author{Giovanni Barontini}
\email{g.barontini@bham.ac.uk}
\affiliation{Midlands Ultracold Atom Research Centre, School of Physics and Astronomy, University of Birmingham, Edgbaston, Birmingham, B15 2TT, United Kingdom}
\author{Mauro Paternostro}
\affiliation{Centre for Theoretical Atomic, Molecular and Optical Physics,
School of Mathematics and Physics, Queen's University Belfast, Belfast BT7 1NN, United Kingdom}

\date{January 24, 2018}

\begin{abstract}
We propose a scheme for a single-atom quantum heat engine based on ultra-cold atom technologies. Building on the high degree of control typical of cold atom systems, we demonstrate that three paradigmatic heat engines -- Carnot, Otto and Diesel -- are within reach of state-of-the-art technology, and their performances can be benchmarked experimentally. We discuss the implementation of these engines using realistic parameters and considering the friction effects that limit the maximum obtainable performances in real-life experiments. We further consider the use of super-adiabatic transformations that allow to extract a finite amount of power keeping maximum (real) efficiency, and consider the energetic cost of running such protocols. 
\end{abstract}


\maketitle 

The role played by thermodynamics in our daily life can hardly be emphasized enough. The heat machines and refrigerators that are widely employed in industry and in transports are essentially based on elementary thermodynamic cycles. On the other hand, as already visionarily predicted by Feynman in 1960 \cite{uno}, technological progress is pushing towards the realization of smaller-scale devices and, at the ultimate level, machines will be built only with one or a few atoms. In such operating regime of low-scale energies, questioning whether the paradigm of thermodynamics needs a fundamental redefinition, including a quantum mechanical formulation of \emph{heat} or \emph{work}, is quite natural. 

It is in this context that the emerging field of Quantum Thermodynamics comes into play, with the aim of including quantum mechanical effects into the thermodynamic framework. There are several theoretical studies that have been put forward, addressing theoretical aspects of such reformulations~\cite{dodici, tredici,quattordici, quindici} and extending all the way to the assessment of quantum engines~\cite{sei,sette,otto,undici}. Quantum engines exploiting a quantum-coherent working fluid have been proven to generate substantially more power than classical stochastic engines~\cite{diciotto}. Moreover, non-thermal (non-classical) baths and many-body effects can lead to more efficient and powerful engines~\cite{venti,dillenschneider,rossnagel,ventidue}, and sophisticated control techniques can be used to enhance such performances even further~\cite{ventuno}.	

Such substantial theoretical advance is yet to be translated into feasible experimental platforms. While recently nitrogen vacancy (NV) centres in diamond and ultra-cold atoms have been used to demonstrate quantum features in the operation of a heat engine~\cite{merdacce,brantut}, to date only one experiment \cite{ventitre} has reported a single-atom engine, although operating fully in the the classical regime. 
In this paper we go beyond such limitations and discuss an architecture based on cold atom technology for the realization of single-atom engines that are able to enter the quantum domain of operation. 
The ultra-cold temperatures that characterize our operating system guarantee that the engine works in a fully quantum regime. We show that using our architecture based on ultracold atomic mixtures, we can arrange for arbitrary thermodynamic transformations and thus, in turn, arbitrary thermodynamic cycles, including the quantum Carnot, Otto and Diesel ones~\cite{sei,sette,otto,nove}. Moreover we design super-adiabatic transformations that allow to reach high efficiencies in finite time and we discuss the friction effects that limit the performances of real ultra-cold atomic engines (UAEs). 

Our UAEs are assembled starting from three basic elements: \textbf{i)} a single ultracold atom that is the working fluid, \textbf{ii)} a species-selective optical tweezer that acts as a piston, \textbf{iii)} a thermal cloud of ultracold atoms of a different species that embodies the thermal bath. The use of two ultra-cold atomic species allows the control of the fluid-bath interaction with an external magnetic field through Feshbach resonances and ``zero-crossings"  of the scattering length. These are used to accurately control and also turn off the interaction between the bath and the system~\cite{bloch,julienne}. Possible implementations include but are not limited to Cs-Rb \cite{grimm}, Cs-K \cite{hutson} Li-Cs \cite{chin,repp} and K-Rb~\cite{due}. Another key ingredient is the use of a species-selective optical tweezer that is transparent for the bath atoms but that allows, at the same time, the \emph{selective} trapping and manipulation of the single atom of the other species~\cite{clark}. {Interestingly, optical tweezers have been similarly used to realize classical micro-engines \cite{tweezers}}. 
The tweezer can be designed so that the transverse trapping frequencies set an energy scale that is much higher than the thermal one corresponding to the operating temperature, while the axial frequency energy scale is comparable with the thermal one. This allows only the population of the lower axial energy levels and the level corresponding to the radial ground state, thus realizing an effective one dimensional multi-level system on the axial degrees of freedom. Thermodynamic transformations on the working fluid are performed by the tweezer-piston and controlling the atom-bath interaction.  The bath is confined in a large scale trap, so that the single atom is not affected by the modifications of the bath trapping potential. {Such architecture can be easily realized using standard cold atoms techniques like evaporative cooling and sympathetic cooling.}


\begin{figure}
	\centering
		\includegraphics[width=0.8\columnwidth]{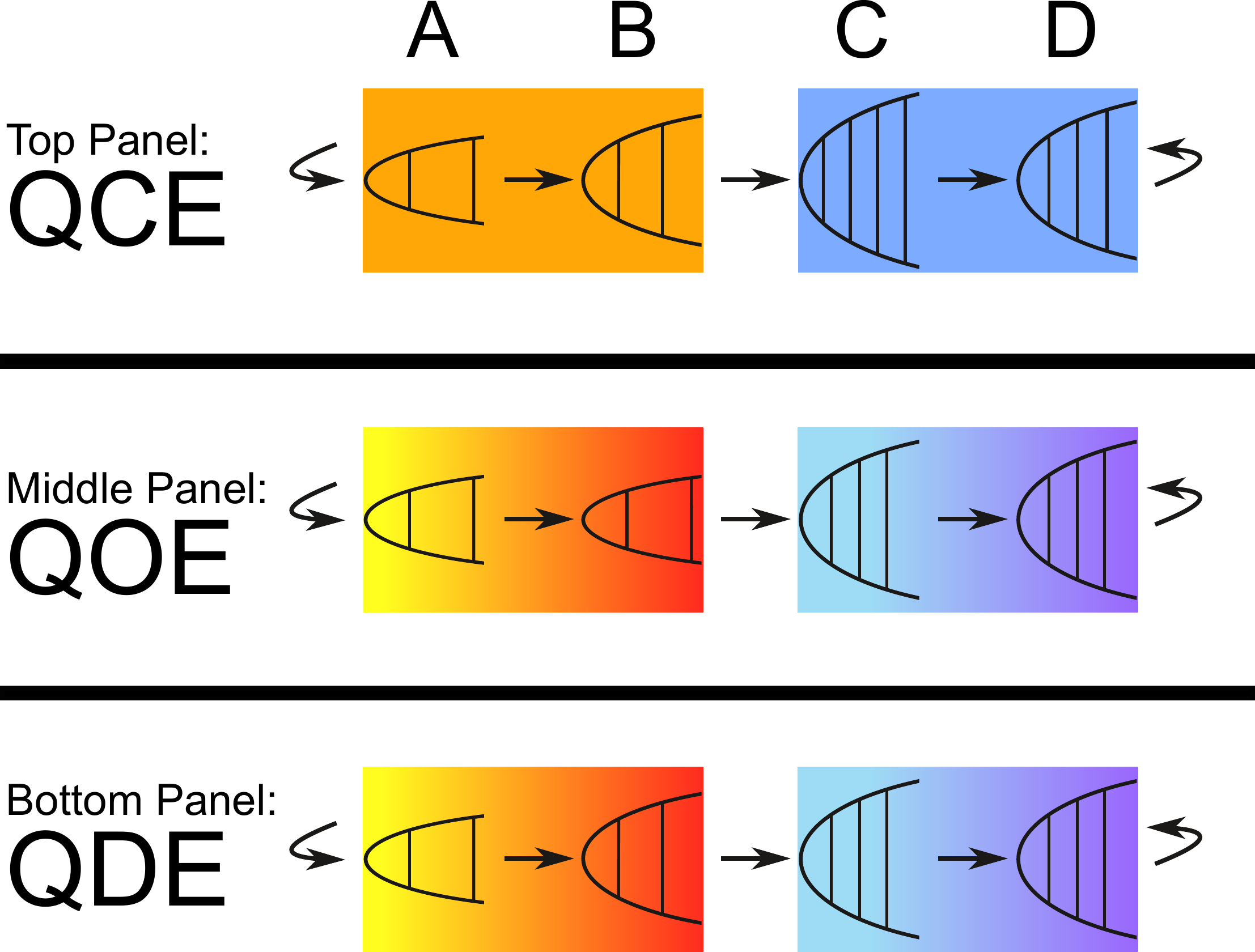}
	\caption{From top to bottom: ultra-cold single-atom realization of a Quantum Carnot Engine (QCE), a Quantum Otto Engine (QOE) and a Quantum Diesel engine (QDE). The frequency of the harmonic confinement of the working fluid is changed in time (by tightening or loosening the trap) in order to realize the compression or expansion of the wavefunction of the single atom. The background colours indicate the bath temperature: colors in the red range stands for a hot bath, while blue ones are for a cold bath. The (B$\rightarrow$C) and (D$\rightarrow$A) transformations are realized by decoupling the working fluid and the bath.}
\label{fig2}
\end{figure}

We can effectively describe the working fluid of the UAE as a one-dimensional multi-level system on the external axial degrees of freedom and consider the radial degrees of freedom as frozen. Therefore, we can write its Hamiltonian as $H=\sum_n E_n|n\rangle\langle n|$ where $E_n<E_{n+1}$ and $|n\rangle$ are the eigenvalues and eigenvectors of the one-dimensional harmonic oscillator. If $P_n$  is the occupation probability of the $n^{\text{th}}$ level, the total energy of the system is $E=\sum P_nE_n$. We have therefore that a modification of the total energy implies
\begin{equation}
dE=\sum dP_nE_n+\sum P_ndE_n,
\end{equation} 
that is a formulation of the first law of thermodynamics at the single-atom quantum level. In analogy with the classical formulation of heat and work, we can identify the heat exchange to be $dQ=\sum_n E_n dP_n$ and the work exchange in the single atom quantum regime to be $dW=\sum_n P_ndE_n$ \cite{sei,sette,otto}. From these definitions we can write the thermal entropy to be $S=-k_B\sum_n P_n\ln P_n$, with $k_B$ the Boltzmann constant, and the quantum pressure to be $\Pi=-dW/dV$, where $V$ is the trapping volume. When the single atom thermalizes with the bath, the probability $P_n$ that the $n^{\text{th}}$ level is occupied follows the Boltzmann distribution $P_n=1/Ze^{-E_n/k_BT}$, with $Z$ the partition function and $k_B$ the Boltzmann constant. Starting from these considerations it is possible to design the following four basic quantum thermodynamic transformations, which are the basis of a UAE.

\emph{1. The quantum adiabatic transformation} requires the decoupling of system and bath. Then the trapping potential must be compressed (released) while satisfying the condition $dS=0$ and therefore $dP_n=0$. In turn, this implies $dQ=0$. This is achieved setting the external magnetic field to the exact zero-crossing of the interspecies scattering length and by increasing (decreasing) the laser power of the tweezer.

\emph{2. The quantum isothermal transformation} requires to switch on the interaction between the single atom and the thermal bath and to compress (decompress) the potential trapping the atom. In this case the single atom absorbs (emits) heat from the bath at constant temperature during the compression (expansion) of the energy levels. 

\emph{3. The quantum isochoric transformation} preserves the volume of the quantum system. Therefore no transformations of the external potential are involved, i.e., $dE_n=0$ (no work done). During these transformations, the system is put in thermal contact with the bath, whose temperature changes in time. This leads to a change in the occupation probability distributions  $dP_n\neq0$ and therefore  $dS\neq0.$ This is realized by putting the single atom in interaction with the thermal bath and changing the temperature of the latter by compressing or decompressing its trapping potential.

\emph{4. The quantum isobaric transformation} keeps the pressure or the force on the working fluid constant. The working fluid is in contact with the bath and the temperature is changed together with the trapping potential. We consider a one dimensional harmonic oscillator with $E_n=\hbar\omega(n+1/2)$, $\omega/2\pi$ being the harmonic oscillator frequency. The pressure in the working fluid can be calculated as $\Pi=-\sum P_n(dE_n/da_{ho})$, with $a_{ho}=\sqrt{\hbar/m\omega}$ the harmonic oscillator length, yielding
\begin{equation}
\label{pressure}
\Pi=\frac{\sqrt{\xi}\sinh\beta}{\cosh\beta-1}
\end{equation}
with $\xi=h m\omega^3$ and $\beta=\hbar\omega/(k_BT)$. From Eq.~(\ref{pressure}) it follows that, during an isobaric transformation, the temperature must be changed according to
\begin{equation}
\frac{\hbar\omega}{k_BT}=\ln\left(\frac{\Pi+\xi^{1/2}}{\Pi-\xi^{1/2}}\right).
\end{equation}

Using UAEs it is therefore possible to implement all the four elementary quantum thermodynamic transformations. For the realization of the quantum engines it is then necessary to combine those transformations in cycles. We first analyse the Quantum Carnot Engine (QCE). Its classical counterpart is the paradigm of every engine and its efficiency sets the maximum theoretical efficiency that any engine (either classical or quantum) can achieve. The QCE is composed of four transformations (cf. the top panel of Fig. 1): {\bf 1)} A hot quantum isothermal expansion at temperature $T_1$ (A$\rightarrow$B) in which the working fluid receives heat from the thermal bath; {\bf 2)} A quantum adiabatic expansion (B$\rightarrow$C) in which work is extracted from the working fluid; {\bf 3)} A cold quantum isothermal compression at temperature $T_2<T_1$ (C$\rightarrow$D) in which the working fluid transfers heat to the thermal bath; {\bf 4)} A quantum adiabatic compression (D$\rightarrow$A) in which work is done on the working fluid. To close the cycle, the change in temperature must fulfil the relation $T_1/T_2= \Delta_B/\Delta_C=\Delta_A/\Delta_D$, where  $\Delta_i=\hbar\omega_i$ \cite{sei}. As for its classical analogous, the maximum efficiency of the QCE is $\eta_{max}=1-T_2/T_1$, which can be achieved in principle only with quasi-static transformations. This implies that, at maximum efficiency, no power can be extracted from the engine. However, one of the advantages of UAEs is the ease to implement super-adiabatic transformations~\cite{superad,superad1}. These allow to follow the self-similar evolution of the initial state, implying that the condition $dP_n=0$ is rigorously fulfilled at every instant of time of the dynamics. Therefore, no friction is produced and power can be extracted in a finite time keeping maximum efficiency. With the UAE we take advantage of the fact that both the working fluid and the bath are confined in harmonic potentials, for which super-adiabatic transformations can be calculated analytically~\cite{ventuno}. By assuming the time-dependence of the harmonic trap frequency, which we label as $\Delta_t$, we have that the ideal controlling process should read
\begin{equation}
\Delta_t=\sqrt{\frac{\Delta^2_0}{b(t)}-\frac{b''(t)}{b(t)}}
\end{equation}
with $b(t)=1+t^{*3}\sqrt{\Delta_0/\Delta_{t_f}}(6t^{*2}-15t^*+10)$, where $t^*=t/t_f$ is a dimensionless evolution time defined with respect to the the duration of the transformation $t_f$.

In order to give a specific example, we study the case of the $^{87}$Rb-$^{41}$K mixture. However, it is very important to remark that, qualitatively, our results do not depend on the choice of mixture. We use the Rb atoms as elements fo the bath, while a single K atom embodies the working fluid. The single K atom is loaded into the tweezer at the exact species-selective wavelength that, for this mixture, is 789.82 nm (assuming linearly polarized light) \cite{tre,lamporesi}. This makes the species-selective light red-detuned with respect to the K transitions, therefore suitable for trapping. The $^{87}$Rb-$^{41}$K mixture features two interspecies Feshbach resonances and a zero of the scattering length at relalively small magnetic fields \cite{due}.

We first discuss the optimal temperature for the starting stage A of the QCE in the upper panel of Fig. 1. A reasonable choice is $T_A=T_1=\hbar\omega_A/2k_B$, so that $P_0+P_1+P_2>0.99$ for the K atom. The QCE includes two isothermal transformations that require collisions between the single trapped atom and the bath. In the ultra-cold regime the heat capacity of the bath is reduced and, to allow thermalization, the atom-bath interaction has to be made stonger as the temperature is decreased \cite{quarantuno}. However, when increasing the interaction strength, the survival probability of the K atom to three-body losses decreases, due to the fact that the inelastic scattering rate scales as $\propto a^4$ with $a$ the interspecies scattering length. The reduced survival probability  limits the efficiency of a real UAE, therefore providing an \emph{effective friction}. 
Based on the study reported in Ref.~\cite{collisions}, in order to grant thermalization about $\simeq$4 collisions would be required. Therefore, the strength of the interactions will have to depend also on the length of the transformation $t_f$. In general, it is desirable to perform fast transformations to extract a large amount of power and avoid spurious effects such as the heating coming from the light of the optical tweezer. The optimal working point thus depend on the trade-off between the needs to perform fast transformations and reducing the effective friction. In Fig.~2 {\bf a)} we report the ratio of the maximum real efficiency $\eta_{real}$, calculated multiplying the theoretical efficiency with the survival probability, and $\eta_{max}$ as a function of the starting temperature, for a KRb UAE~\cite{SM}. Such efficiencies were calculated considering the length of each isothermal transformation to be 1 ms. Clearly, our UAEs do not allow high real efficiencies for temperatures below 1 $\mu$K. In Fig.~2 {\bf b)} we report $\eta_{real}$ versus $\eta_{max}$ for different lengths of the isothermal strokes $t_f$ and $T_1=1.1 \mu$K. As expected, as the length of the strokes is increased the effective friction decreases. However, if we consider also the photon scattering rate coming from the tweezer, which might induce spurious heating, we find that the optimal starting temperature is $T_1=1.1 \mu$K and that it is convenient to set $t_f=1$ms for each isothermal transformation. This indeed guarantees that the atom-bath scattering rate (4 kHz) is one order of magnitude higher than the maximum photon scattering rate coming from the tweezer ($\simeq$400 Hz) \cite{SM} and boosts power extraction. 
 
\begin{figure}
	\centering
		\includegraphics[width=0.4\textwidth]{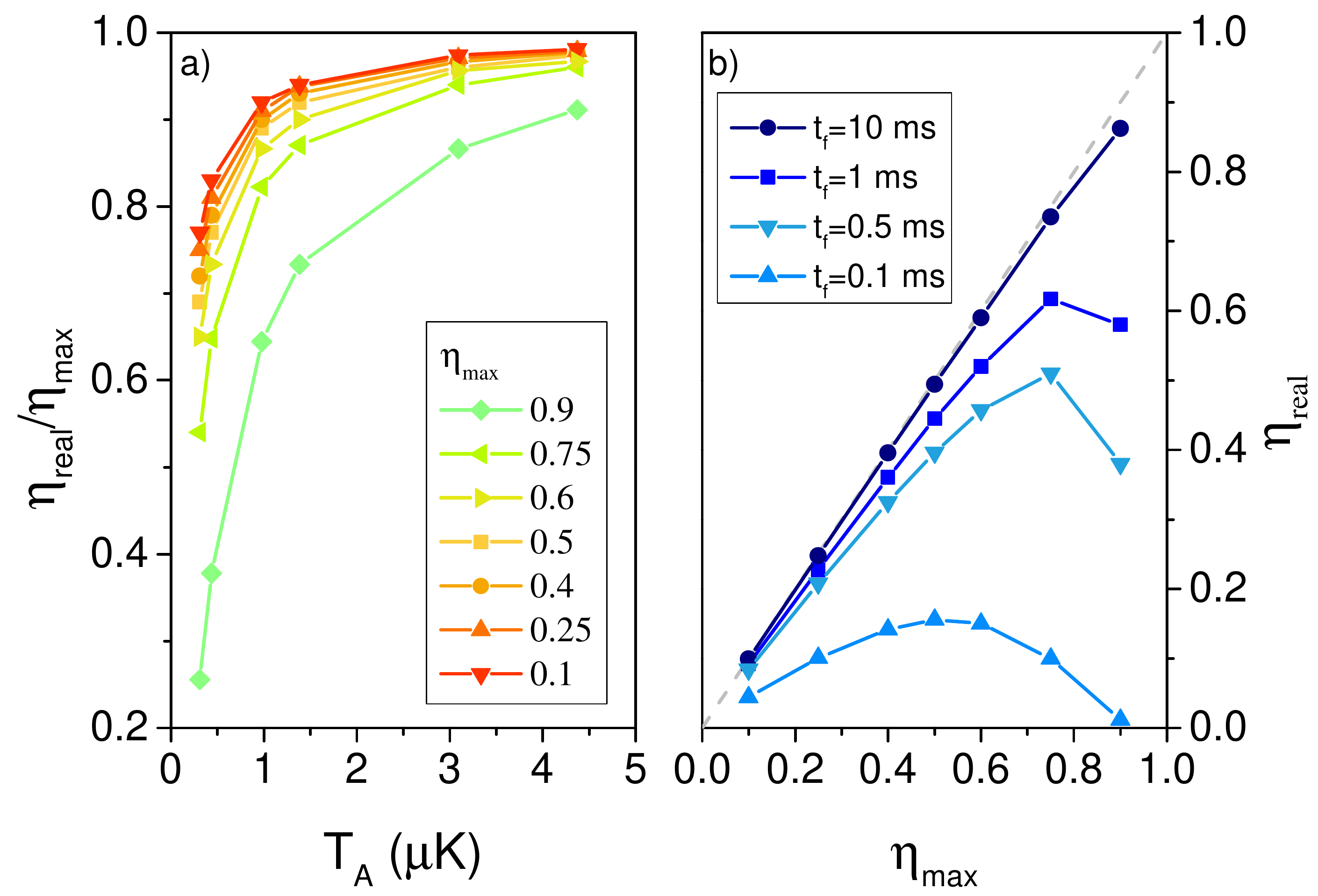}
	\caption{Panel {\bf a)}: Maximum real efficiencies of a KRb UAE as a function of the starting temperature $T_A=T_1$ for a QCE with two isothermal strokes of 1 ms each. Panel {\bf b)}: Maximum real efficiency versus maximum theoretical efficiency for a QCE with $T_1=1.1\mu$K and different lengths $t_f$ of the isothermal strokes.}
\label{fig1}
\end{figure}

After having set the initial conditions, we now focus on the realization of a super-adiabatic QCE with $\eta_{max}=0.75$, so that $\eta_{real}=0.62$, which would be similar to the typical efficiency of a car engine. We thus set $T_2/T_1=0.25$. Similarly, we set the first isothermal expansion factor $\Delta_B/\Delta_A$ to 0.5. Although for the single atom it is possible to achieve super-adiabatic transformations as fast as a few $\mu$s, the speed of the super-adiabatic strokes is set by the maximum speed achievable by the super-adiabatic transformations performed on the bath. Indeed, the temperature of the bath must be changed so that $(\Delta_C/\Delta_B)_{bath}=(\Delta_D/\Delta_A)_{bath}=T_2/T_1$, in parallel with the modification of the trapping potential of the working fluid. In the configuration chosen here, the shortest transformation lasts 0.55 ms, as shown in Fig.~3 {\bf a)}~\cite{SM}. The work through a QCE is $W_{QCE}=(T_1-T_2)(S_B-S_A)$, as for the classical counterpart. For the QCE engine reported in Fig.~3 {\bf a)}, the total cycle time is $\tau=$2.46 ms so that the extracted power is $\mathcal{P}/k_B=W_{QCE}/k_B\tau=0.14$ mK/s, obtained maintaining the maximum real efficiency $\eta_{real}$. {The quantification of the performance of our UAEs can be done using well established techniques. The measurement of the level population of the K atom $P_n$ can be inferred by using Raman sideband spectroscopy \cite{venticinque}, while the temperature of the Rb bath can be obtained with standard time-of-flight imaging. The energy spacing $E_n$ is given by the tweezers parameters. With this diagnostics, it is possible to access all the observables necessary to evaluate the quantum thermodynamic quantities of interest \cite{sei,sette}. It is worth noticing that such measurements are destructive and need to be done only to demonstrate the proof-of-principle. Once the working principle is demonstrated, the functioning of the engine will not need any measurement to be performed.}

\begin{figure}
	\centering
		\includegraphics[width=0.45\textwidth]{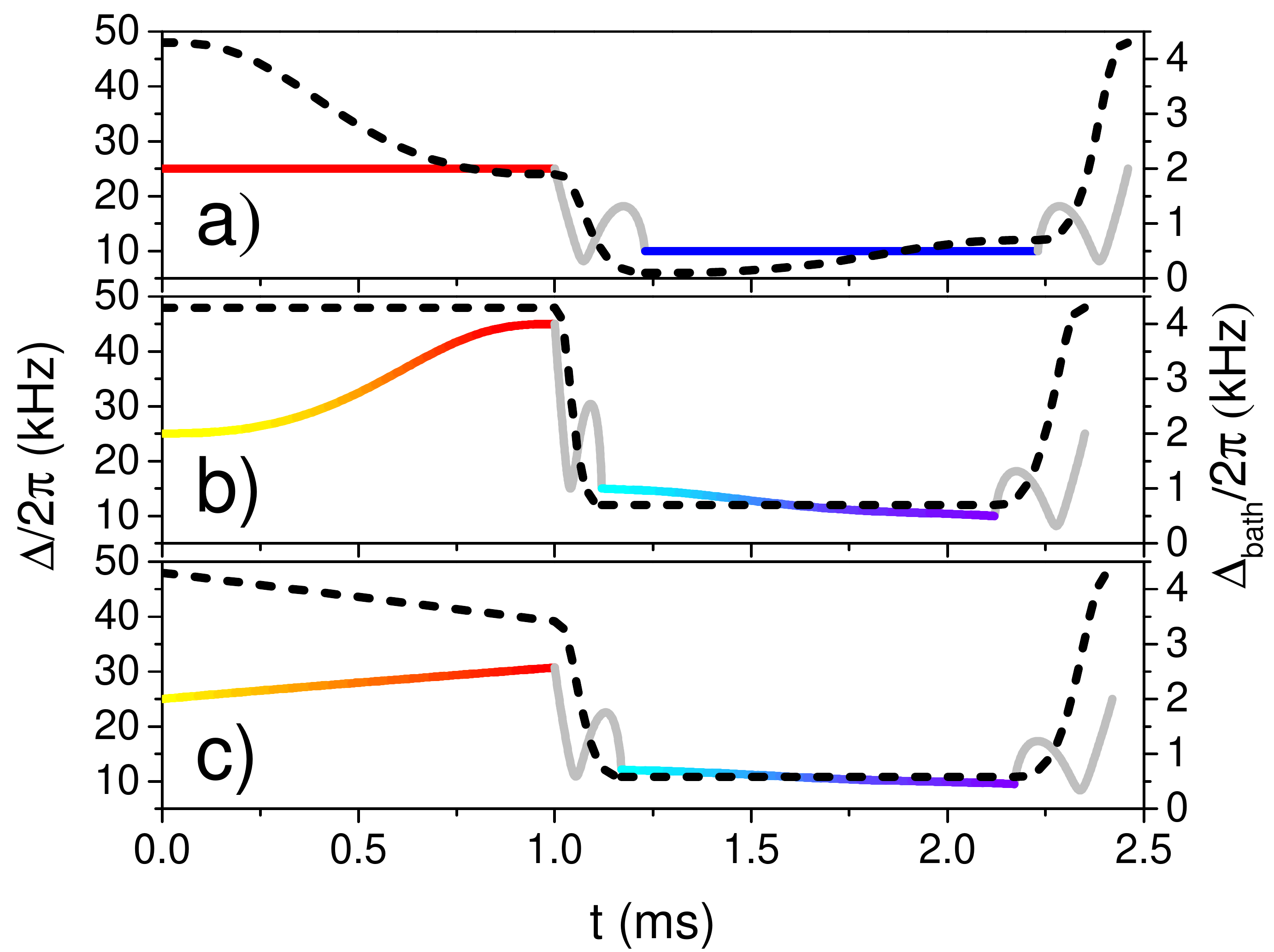}
	\caption{Panel {\bf a)}: Ultra-cold QCE with super-adiabatic strokes. The dotted line is the evolution of the trapping frequency of the working fluid across the cycle. The solid line is the trapping frequency of the bath, that sets the temperatures. The red colour indicates the hot bath at temperature $T_1$ while the blue color the cold bath at temperature $T_2$. During the super-adiabatic transformations there is no contact between the bath and the working fluid (grey color). Panel {\bf b)}: Ultra-cold QOE. Solid and dotted lines have the same meaning as in {\bf a)}. The color scales indicate the change in temperature of the bath during the isochoric transformations. Panel {\bf c)}: Same as {\bf a)} and {\bf b)} but for the QDE.}
\label{fig3}
\end{figure}

\noindent
{\it Quantum Otto engine.--} The second quantum engine that can be implemented with the UAE architecture is the Quantum Otto engine (QOE). Its classical counterpart is the most employed engine in automotive industry. Its working principle is shown in the middle panel of Fig.~1 and described in Ref.~\cite{SM}. To make a direct comparison with the QCE described above, we use the same initial conditions and set $\eta_{max}=0.75$ and $T_B=2T_A$. The power extracted with the QOE shown in Fig.~3 {\bf b)} is $\mathcal{P}/k_B=2.36$ mK/s, with $\tau=2.35$ ms and $\eta_{real}=0.68$ \cite{SM}. Therefore a real ultra-cold QOE is both more efficient and more powerful than a real QCE with the same initial conditions and the same maximum theoretical efficiency.       

\noindent
{\it Quantum Diesel engine.--} The last engine that we take into account in this work is the Quantum Diesel Engine (QDE), which is shown in the bottom panel of Fig. 1. In this case too, in order to make a direct comparison with the processes addressed previously, we choose the same initial condition and $\eta_{max}=0.75$ [cf. Ref.~\cite{SM} for details]. For the optimized cycle in Fig. 3 {\bf c)}, the total cycle time is $\tau=2.42$ ms so we obtain that the power that can be extracted is $\mathcal{P}/k_B=2.54$ mK/s, slightly higher than the QOE. The real efficiency is $\eta_{real}=0.64$, in between the efficiencies of the QCE and the QDE. 

\noindent
{\it Energetic cost of super-adiabaticity.--} A very informative figure of merit to quantify the performance of an engine is the efficiency at maximum power. The approach usually taken when evaluating such parameter is to embed the time dependence of heat transfer in the analysis of the engine. In some cases such as a QOE, this  leads to the assumption of constant finite cycle time, which implies that power and work are treated on the same footing and therefore to the definition of the celebrated Culz\'orn-Alhborn efficiency~\cite{CA}. When considering time-dependent transformations, such as those at the basis of a super-adiabatic approach, other constraints should be included in order to estimate a more faithful indicator of efficiency at optimal values of power. For our UAEs, an important constraint to impose on the energy that is put into the working medium is that no inversion of the harmonic trap of the system should be in order. For simplicity, we assume that no residual non-adiabatic excitations remains at the end of the super-adiabatic protocol. The efficiency at maximum power can evaluated through the expression
\begin{equation}
\eta^*=1-\frac{\gamma+\sqrt{4\gamma(1+\gamma)}}{2+\gamma}
\end{equation}
with $\gamma$ the ratio of mean energy of the working medium at the start of the isentropic compression and expansion, respectively. The explicit evaluation of this quantity for our QOE shows that $\eta^*\simeq80\%$ of the Culz\'orn-Alhborn efficiency, proving that the use of super-adiabatic approaches for the operation of a QOE is effective in delivering high efficiency cycles associated with maximum possible power~\cite{Abah}. 

\noindent
{\it Conclusions.--} We have presented an ultra-cold atom system in which the fundamental thermodynamic transformations can be realized at the quantum level. We have shown how to practically implement the QCE, QOE and QDE, provided a detailed example that takes into account friction effects with the bath and discussed how to engineer super-adiabatic transformations. Our work provides a first step towards the concrete realization of quantum heat engines in the ultra-cold regime, that might give useful insights on the relation between thermodynamics and quantum mechanics and lead to applications in quantum information. {Additionally, the work produced by our UAEs can be extracted and transformed into transport, as shown in Ref.~\cite{SM}. The low temperatures,  together with the finite-time of operation that we have considered, imply that the working medium of our engine is, in general, in a state that is not necessarily thermal. The characterisation of the behaviour of quantum coherence (which can be done following the lines in Ref.~\cite{Baumgratz}) during the operation of our engines, and the establishment of a causal relation with the efficiency of such devices~\cite{Dag, tukpence}, will be the topic of our further investigations.  
Notably, the proposed architecture can be extended to arrays of atoms, allowing to investigate the role of entanglement in quantum thermodynamics. 

\noindent
{\it Acknowledgements.--} The authors are supported by the Leverhulme Trust Research Project Grant UltraQuTe (grant number RGP-2018-266) and acknowledge fruitful discussions with the members of the Cold Atoms Group at the University of Birmingham and Obinna Abah at Queen's University Belfast. MP acknowledges support from the DfE-SFI Investigator Programme (Grant No. 15/IA/2864), and the H2020 Collaborative Project TEQ (Grant Agreement No. 766900).

\newpage

\section{Supplementary materials}

\subsection*{Effective Friction}

We have considered a realistic bath with $10^6$ atoms with average trapping frequency (at stage A) of $2\pi\times$2 kHz. The results are however almost independent on the Rb trapping frequencies. 

To calculate the thermalization rate, we consider the interspecies elastic collision rate in a Krb mixture is \cite{quarantuno}
$$\Gamma=\frac{\sigma_{KRb}vN_KN_{Rb}}{\pi^{3/2}\sum_i\sqrt{\sigma_{iK}+\sigma_{iRb}}},$$
where $\sigma_{KRb}=4\pi a_{KRb}^2$, with $a_{KRb}$ the interspecies scattering length, and $\sigma_i$ with $i=(x,y,z)$ the widths of the atomic distributions. For every stroke, we have imposed the condition $\Gamma=4/t_f$ to allow thermalization. 

The interspecies inelastic event collision rate, i.e., the number of inelastic collision events per unit of time and volume is instead is $\alpha=4\pi\hbar a_{KRb}^4/m$ \cite{sch}. This is responsible for the effective friction caused by three-body losses. For every stroke that needs the contact with the thermal bath we solve the equation for the survival probability in the optical tweezer against three-body losses
$$\frac{d}{dt}n_K(t)=-\alpha n_K(t)n_{Rb}^2(t),$$
with $n_{Rb}$ the density of the Rb atoms in the bath, that leads to an exponential decay.  

The real efficiency $\eta_{real}$ for UAEs is calculated multiplying the maximum theoretical efficiency $\eta_{max}$ with the total survival probability after a complete cycle.

\subsection*{Tweezer heating rate}

To obtain $\omega=2k_BT_1/\hbar=2\pi\times48$ kHz, we employ a tweezer with 1 $\mu$m waist and 7 mW of power, yielding radial trapping frequencies of $\simeq225$ kHz. The maximum photon scattering rate, that scales linearly with the trapping power and therefore with the temperature of operation, is calculated according to these figures \cite{grimm}. For the Rb cloud the heating rate is strongly reduced due to the small volume of the tweezer with respect to the size of the cloud

\begin{figure}
	\centering
		\includegraphics[width=0.5\textwidth]{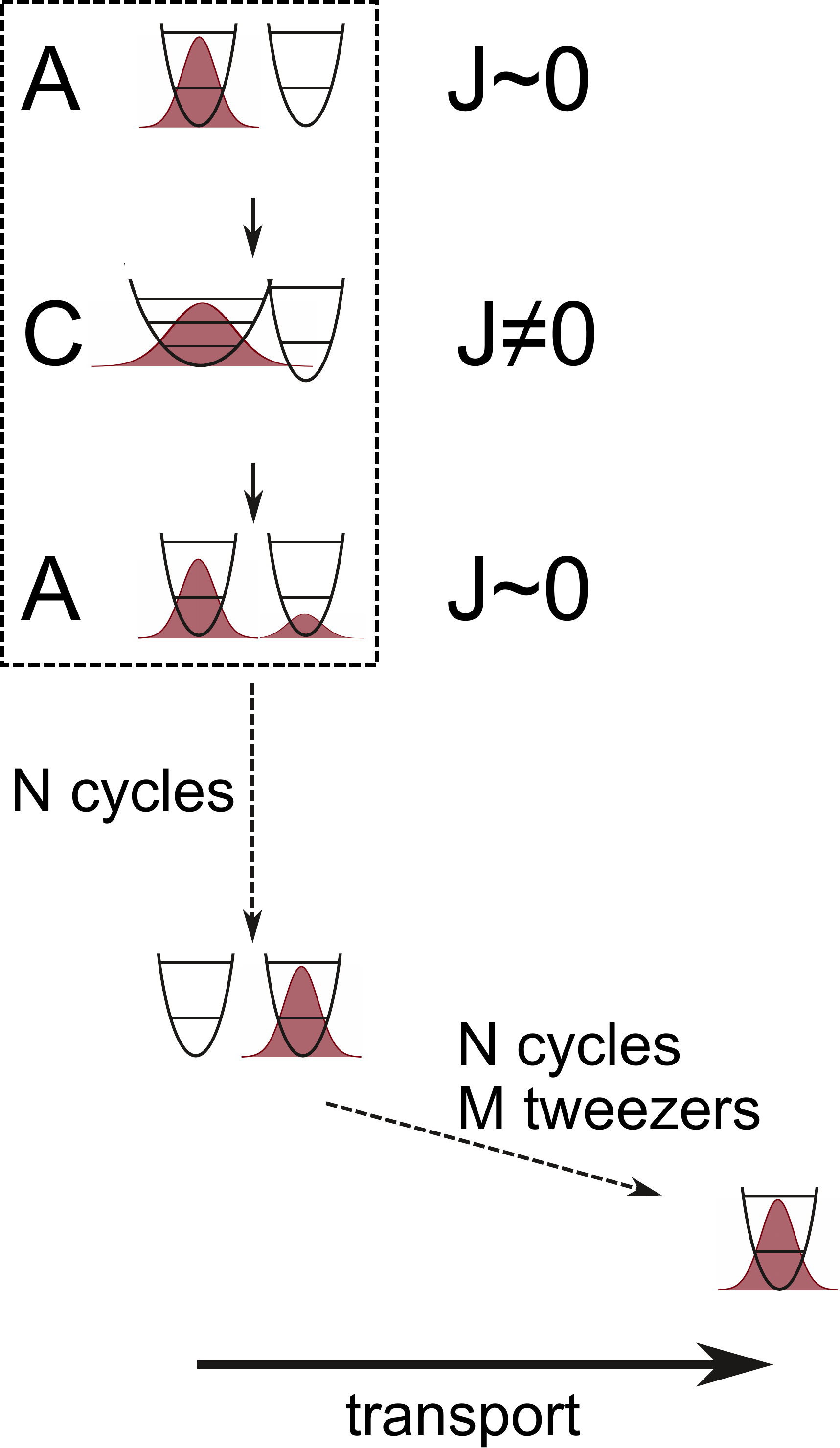}
	\caption{(Color online) Schematic representation of the protocol described in the text that allows to extract the work produced by the UAEs and transform it into motion.}
\end{figure}

\subsection*{Quantum Otto Engine}

The QOE consists in four elementary strokes. With respect to the QCE the quantum isothermal transformations (A$\rightarrow$B) and (C$\rightarrow$D) are replaced by quantum isochoric transformations during which no work is done but heat is absorbed from or released to the bath, see the middle panel in Fig. 1 of the main text. To close the cycle it must be ensured that $T_B/T_D> \Delta_A/\Delta_D$ \cite{sei}. The maximum theoretical efficiency of the QOE is $\eta_{max}=1-\Delta_D/\Delta_A=1-T_C/T_B$ and also in this case it can be reached in finite time using super-adiabatic transformations. To make a direct comparison with the QCE described in the main text, we use the same initial conditions and impose $\eta_{max}=0.75$ and $T_B=2T_A$. In the QOE we employ super-adiabatic transformations on the bath also during the isochoric strokes to minimize friction effects in the bath. With arguments similar to those detailed for the QCE, to guarantee thermalization and avoid excessive friction we choose the length of these strokes to be 1 ms. The super-adiabatic transformations are similar to those obtained for the QCE, although the (B$\rightarrow$C) one can be slightly faster due to the reduced decompression, see Fig.3b in the main text. Also the effective friction of the QOE is reduced with respect to the QCE due to the higher average temperature of the two baths, indeed  $\eta_{real}=0.68$. The work done by the QOE in one cycle is
$$W_{QOE}=\sum[E_n(B)-E_n(C)]\sum[P_n(B)-P_n(A)],$$
hence the power extracted is $\mathcal{P}/k_B=2.36$ mK/s, with $\tau=2.35$ ms.

\subsection*{Quantum Diesel Engine}      

The QDE is again made by four transformations: the (B$\rightarrow$C), (C$\rightarrow$D) and (D$\rightarrow$A) are the same as those of the QOE but the (A$\rightarrow$B) isochoric transormation is replaced by a quantum isobaric expansion in which heat is transferred to the working fluid, see the bottom panel of Fig. 1 in the main text. The efficiency of the QDE is given by 
$$\eta_{max}=1-(\Delta_D/\Delta_A)[(\Delta_A/\Delta_B)^{3/2}-1]/[3(\Delta_A/\Delta_B)^{1/2}-3].$$ 
Also in this case, to make a direct comparison we choose the same initial condition and $\eta_{max}=0.75$. We additionally impose that $\Delta_D/\Delta_A$ is 0.9 times the value  in the QOE, that yields to $\Delta_A/\Delta_B\simeq1.23$. During the isobaric stroke the temperature of the bath must be changed carefully according to eq. (3), for simplicity we use a linear compression ramp. No super-adiabatic transformations can be used in the bath during this stroke. Choosing again 1 ms for the strokes that require thermalization with the bath, we find that the real maximum efficiency for the QDE is $\eta_{real}=0.64$, in between the efficiencies of the QCE and the QDE. The cycle is shown in Fig. 3a in the main text. The work done by the QDE is 
$$W_{QDE}=\Pi_{AB}\Delta V_{AB}-\sum E_n(C)[P_n(C)-P_n(D)],$$ 
where $\Delta V$ is the change in volume during the isobaric expansion. The total cycle time is $\tau=2.42$ ms so we obtain that the power that can be extracted is $\mathcal{P}/k_B=2.54$ mK/s.

{
\subsection*{Work Extraction}
We now describe an experimental scheme that allows to transform the work produced by our UAEs into transport. The scheme relies on the use of an array of tweezers, as for example those in \cite{endres, barredo} and is depicted in Fig. 1 of this supplementary materials. Let us first consider two identical tweezers set at a certain distance so that the tunnelling between them $J$ is suppressed. A single atom is loaded in one of them (say the left one) and is prepared in the configuration corresponding to stage A of our engines. We then start the engine and when stage C is achieved the atomic wavefunction has expanded sufficiently to allow the tunnelling into the right tweezer. This means that after a complete cycle we will have that in the left tweezer $\sum_n P_n<1$. After $N$ cycles the atom will be completely transferred into the right tweezer. This process can be repeated for an arbitrary number of tweezers allowing to use the work produced by the atom to transport it over an arbitrary distance. The quantitative study of this scheme will be addressed in a future work.   
}

\end{document}